\documentclass[a4paper]{article}
\usepackage{spconf,amsmath,graphicx,epsfig}
\usepackage{subfigure}
\usepackage{color}
\usepackage[table]{xcolor}
\usepackage{multirow}
\usepackage[table]{xcolor}
\usepackage{float}
\usepackage{url}
\usepackage{soul}
\title{Data Augmentation Enhanced Speaker Enrollment for Text-dependent Speaker Verification}
\name{Achintya Kumar Sarkar$^1$, Himangshu Sarma$^1$, Priyanka Dwivedi$^1$,  Zheng-Hua Tan$^2$}
\address{$^{1}$ Indian Institute of Information Technology (IIIT), Sri City, Chittoor, India\\
$^{2}$Department of Electronic Systems, Aalborg University, Aalborg, Denmark\\
{\small \tt sarkar.achintya@iiits.in}
}

\begin{document}
\ninept
\maketitle
\begin{abstract}
Data augmentation is commonly used for generating additional data from the available training data to achieve a robust estimation of the  parameters of complex models like the one for speaker verification (SV), especially for under-resourced applications. SV involves training speaker-independent (SI) models and  speaker-dependent models where speakers are represented by models derived from an SI model using the training data for the particular speaker during the enrollment phase.  While data augmentation for training SI models is well studied, data augmentation for speaker enrollment is rarely explored. In this paper, we propose the use of data augmentation methods for generating extra data to empower speaker enrollment.
Each data augmentation method generates a new data set. Two strategies of using the data sets are explored: the first one is to training separate systems and fuses them at the score level and the other is to conduct multi-conditional training. 
Furthermore, we study the effect of data augmentation under noisy conditions. 
Experiments are performed on RedDots challenge 2016 database, and the results validate the effectiveness of the proposed methods. 

\end{abstract}


\section{Introduction}
Speaker verification (SV) \cite{kinnunen2010overview} is defined as the task of verifying a person using their voice signal.
It is a binary classification problem, where an SV system takes decision by either accepting or rejecting a person claiming the identity using his/her voice.  
As in most of machine learning methods, constructing an SV system consists of training and test phases. In the training/enrollment phase, speakers are characterized by their models/vectorized representation using his/her speech samples during training. In test, a speaker requests to grant the access of a system by claiming his/her identity with a voice sample. The delivered (test) speech sample is then scored against the claimant specific speaker representation in the system. Finally, the score is used for decision making whether  the claimant will be accepted or rejected.

SV systems can be broadly divided into text-independent (TI) and text-dependent (TD). In TI-SV, speakers are free to speak any sentences during the enrollment and verification processes, whereas TD-SV constraints a speaker to speak a particular sentence during both the enrollment and  verification/test phases. Since TD-SV maintains the matched phonetic contents between the enrollment and verification phases in contrast to the TI-SV, TD-SV yields lower error rates in speaker verification using short speech utterances.  Therefore,  TD-SV is suitable for real-time applications compared to TI-SV and is the focus of this paper. 

There are many techniques available in the literature for the improvement of TD-SV systems using short utterances. Those techniques can be divided into different domains. For example, \emph{feature-domain} approaches include Mel-frequency cepstral coefficients (MFCC) \cite{Davis80}, perceptual linear prediction (PLP) \cite{Hermansky90}, deep neural networks (DNNs) based Bottleneck feature \cite{journals/taslp/SarkarTTSG19}, while \emph{model domain} methods include Gaussian mixture models-universal background model (GMM-UBM) \cite{reynold00}, i-vector \cite{Deka_ieee2011} and x-vector \cite{conf/icassp/SnyderGSPK18} techniques.

In low resource applications, it is difficult to get a large amount of diverse data  for training a large number of parameters in speaker independent model like GMM-UBM, DNNs, i-vector and x-vector. 
To create the diverse version of available training data, many augmentation techniques have been introduced in literature. Augmentation basically generates  additional data from existing data with sort of transformations, for example, vocal tract length perturbation \cite{Jaitly_vocaltract}, mixing noise or other speech files with the given raw speech signal \cite{noise_mix,Lasseck2018}, applying impulse (IR) response (of hall room, class room) on the given raw speech signal \cite{Stewart2010}, quadratic distortion on raw audio signal (harmonic distortion) \cite{Mauch2013}, wow re-sampling \cite{Mauch2013}, pitch shifting \cite{7829341}, SpecAugment (deformation of log mel spectrogram with frequency masking) \cite{Park_2019} and random image warping \cite{24792} on image.  
The effectiveness of data augmentation has been proven in various studies including speech recognition \cite{Park_2019}, speaker recognition \cite{conf/icassp/SnyderGSPK18}  and image processing \cite{Shorten2019}. 

In speaker verification, augmented data, e.g. noisy version of available training data, are conventionally used to build speaker-independent (SI) modeling, e.g. GMM-UBM \cite{I4U2013, michelsanti2017conditional}, DNNs \cite{conf/icassp/SnyderGSPK18}, total variability space in i-vector \cite{conf/icassp/SnyderGSPK18}, and in post-processing/scoring step e.g. probabilistic linear discriminate analysis (PLDA) \cite{conf/icassp/SnyderGSPK18,achintya2012}. In \cite{7528399,I4U2013}, the noisy version of training speech utterances/speaker enrollment data has been included in the enrollment phase for building a noise-robust model for spoofing detection \cite{7528399} and speaker recognition \cite{I4U2013} under noisy environments, respectively. 
However, as per our best knowledge, there is no study in the literature to  
use augmented data for \emph{speaker enrollment}, other than creating a noisy version of speech data for the purpose of noise robustness. Therefore, it is interesting to investigate whether the class of data augmentation methods including pitch shifting, harmonic distortion,  impulse response and mixing speech file  are useful for speaker enrollment in TD-SV.

\begin{figure*}[ht]
\centering\includegraphics[width=15.4cm,height=10.0cm]{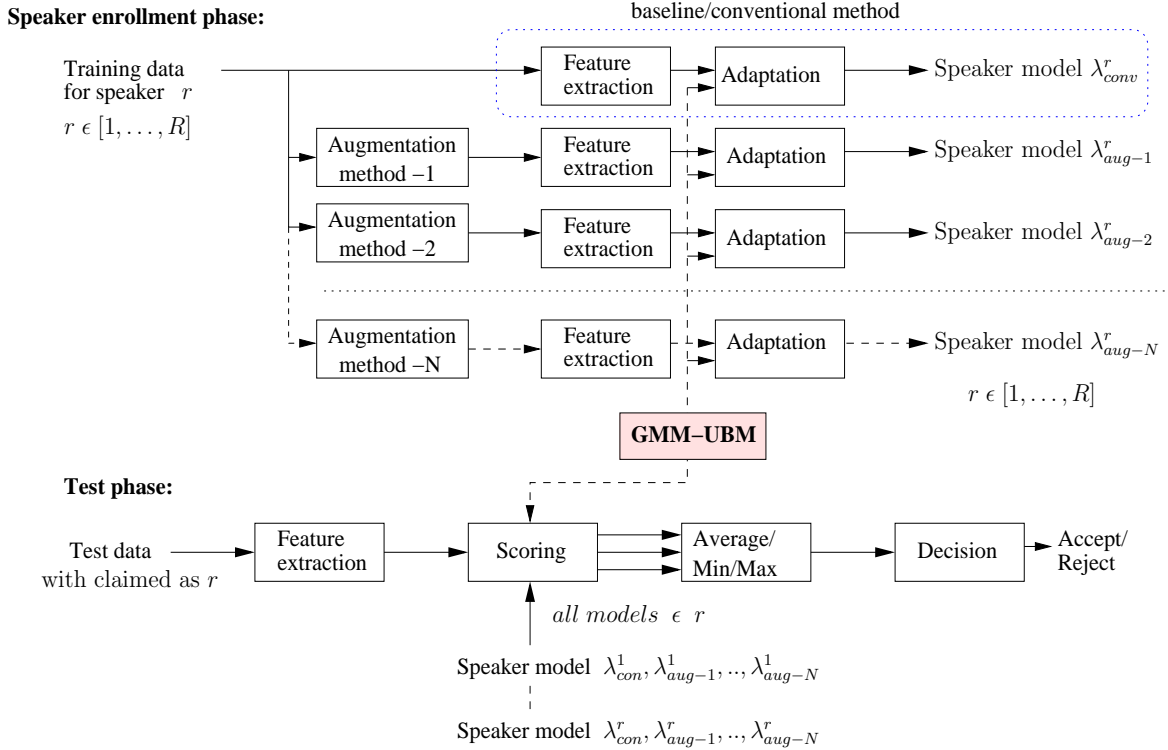} 
\caption{\it Text-dependent speaker verification using augmentation and original training data}
\label{fig:GMM-UBM}
\end{figure*}

The main goal of this paper is to study the effect of different data augmentation techniques to increase the quantity of speaker enrollment data on the performance of speaker verification.  We consider different strategies. First, speaker dependent models are trained for the particular augmentation method in the training phase and in the test phase, original evaluation data without augmentation are scored against the respective speaker models i.e. claimant specific models. It basically develops a separate SV system for each augmentation method.
Score for a given test utterance from different systems are fused into a single value with average, maximum, minimum and median operations. Next, we also study the multi-conditioning training, where a speaker model is trained by pooling both augmentation data generated from different augmentation methods along with the original enrollment data in the database. The performance of TD-SV is studied on the RedDots challenge 2016 database \cite{RedDots} with the GMM-UBM technique. It is well know fact \cite{Delgado2016Asru} that GMM-UBM yields lower error rate for speaker recognition using short utterances than the i-vector technique. Experimental results show that  data augmentation reduces the error rates of TD-SV.

The paper is organized as follows: Section \ref{sec:aug_technique} describes the data augmentation methods. 
Section \ref{sec:GMM_tech} describes the GMM-UBM technique for TD speaker verification.  Experimental setup, and  results and discussion are presented in Section \ref{sec:expsetup} \& Section \ref{sec:resdis}, respectively. Finally, the paper is concluded in Section \ref{sec:con}.

\section{Data augmentation methods}
\label{sec:aug_technique}
In this section, we  briefly describe the different data augmentation techniques considered for TD-SV in this paper. 
\begin{itemize}
    
     \item Pitch shift \cite{7829341}: In this method, the frequency of the voice/speech signal is either increased or decreased without affecting it's duration. We consider two values, namely $\left\{ 1, 2 \right\}$ in semitones, for pitch shifting of a given speech file in speaker's enrollment data.
     \item  Wow re-sampling \cite{Mauch2013}:  This method is similar to the pitch shifting except for changing the  intensity  of the speech signal $x$ along the time,
      \begin{eqnarray}
            \phi(x) = x + a \frac{\sin(2\pi fx)}{2\pi f}
      \end{eqnarray}
      where $\phi(x)$ is the transformed signal. $a$ and $f$ are the control parameters. We consider the change of minimum and maximum intensity and frequency  values up to $3$ and $2$, respectively.
     \item Harmonic distortion \cite{Mauch2013}: It degrades the speech signal by applying   $\sin(.)$ function on it multiple times. The value of degradation  factor is considered to be $5$ as per default parameters available in the toolbox.
     \item Impulse response (hall room) \cite{Stewart2010}: It modifies the given speech signal passing through a simulated source function/filter, e.g. acoustic impulse function/response of a class room. It can be thought as passing the speech signal though a filter which changes the input signal as per the characteristic of the responsive system. 
     
     \item Sound mix \cite{Lasseck2018}: It generates the modified speech signal by adding other audio files from \emph{within the same speaker}. The generated speech will contain the attributes belonging to the same class. However, generated speech could be like as babble noise  due to the overlapped of same person voice.
    \end{itemize}
More details on augmentation techniques can be found in \cite{Mauch2013}.

\section{GMM-UBM technique}
\label{sec:GMM_tech}
In this approach, a larger Gaussian mixture models (GMMs)  \cite{reynold00} is trained using data from many non-target speakers \emph{called GMM-UBM}. The GMM-UBM represents a large acoustic model space which covers the various attributes available in the data. In the enrollment phase, speaker dependent models (of the registered speakers)   are then derived from the GMM-UBM using the training/enrollment data for the particular speaker with maximum a posteriori (MAP) adaptation. In the test phase, the feature vectors of the test utterance $\mathbf{X}=\left\{\mathbf{x_1},\mathbf{x_2},\ldots, \mathbf{x_T} \right\}$ is aligned against the claimant $\lambda_r$ (obtained in enrollment phase) and GMM-UBM $\lambda_{ubm}$ models, respectively. Finally, a log-likelihood ratio value $\Lambda(\mathbf{X})$ is calculated using scores between the claimant and GMM-UBM models and is used to decide whether the claimant will be accepted or rejected.

\begin{eqnarray}
\Lambda(\mathbf{X}) = \frac{1}{T} \sum_{t=1}^{T} \Big[ log\; p(\mathbf{x_t}|\lambda_r) - log \;p (\mathbf{x_t}|\lambda_{ubm}) \Big]
\end{eqnarray}

Fig.\ref{fig:GMM-UBM} illustrates the TD speaker verification system using augmentation and original enrollment data (available in the database), where speaker models are trained using augmentation (generated from the speaker's available enrollment data with augmentation methods) and original enrollment data, separately. Hence, it develops a number of TD-SV systems depending on the use of different enrollment data. In the test phase, scores of the test utterance from different TD-SV systems are fused with \emph{average/minimum/maximum} operations shown in 
\emph{average/min/max}  block,  respectively.

\begin{table}[ht!]
\caption{\it Number of trials available in RedDots evaluation condition for  \emph{m\_part\_01} task.}
\begin{center}
\begin{tabular}{|l|l|l|l|}\cline{1-4}
\# of  & \multicolumn{3}{c|}{\# of trials in Non-target type} \\ 
  Genuine            & Target  & Imposter & Imposter \\
  trials             &-wrong   &  -correct &  -wrong    \\   \hline
       3242       & 29178       & 120086            & 1080774  \\ \hline
\end{tabular}
\end{center}
\label{table:no_trial}
\end{table}

\begin{table*}[!ht]
\caption{\it Comparison performance TD-SV for different  enrollment data and fusion strategy  on RedDots database (m-part01 task).}
\vspace*{+0.2cm}
\begin{center}
\begin{tabular}{llccll|l}\cline{1-7}
System   & Speaker    & Evaluation   & \multicolumn{3}{c|}{Non-target type [\%EER/(MinDCF$\times$ 100)]}  &  Average\\
         & Enrollment &   data       & target-wrong      & imposter-correct & imposter-wrong & EER/MinDCF \\ \hline 
a  (Baseline) &  Original & Original & 3.96/1.54      & 2.79/1.33& 0.92/0.25 & 2.55/1.04            \\
 b &  Wow resampling    & " & {\bf 3.65}/1.56 & 2.95/1.39 & 0.98/0.26 & {\bf 2.53}/1.07 \\
 c &  Pitch shift       &  " &14.33/5.14 & 12.32/5.46 & 8.66/2.76 & 11.77/4.45 \\
 d &  Harmonic distort  &  " & 14.68/4.97 & 11.48/4.33 & 8.36/2.56 & 11.51/3.95 \\
 e & IR hall room      &  " & 26.50/8.34 & 22.39/8.05 & 19.89/6.63 & 22.93/7.67 \\
 f &  Sound mix         &  " & 8.07/3.01 & 6.45/2.87 & 3.96/1.17 & 6.16/2.35 \\ \hline
 {\bf Score fusion}          & {\bf Method}      &          &           &           &             \\
  Systems (a-f)         & Average     &   &  4.53/1.88 &3.39/1.69 &1.41/0.42  &3.11/1.33 \\
                       &  Minimum    &   & 16.03/5.96 & 11.25/5.18& 9.37/3.46  &12.22/4.87 \\
                  &  Maximum    &   & 3.60/1.52 &2.96/1.40 &0.95/0.35 & {\bf 2.50}/1.09 \\
                       &  Median     &   & 6.90/2.65 & 5.12/2.49 & 2.56/0.78 & 4.86/1.98  \\ \cline{2-7}
 (a,b,f)    & Maximum     &   & {\bf 3.67/1.53} & 2.83/1.33 & {\bf 0.89/0.24}  & {\bf 2.46/1.04} \\ \hline 
 Multi-condition   & (a,b,f) & Original  & 4.34/1.68& 3.08/1.44& 1.60/0.43&3.01/1.18  \\ \hline           
                     
\end{tabular}
\end{center}
\label{table:GMM-res-reddots}
\end{table*}

\begin{figure*}[!ht]
\hspace*{+0.1cm}
\includegraphics[width=12.4cm,height=7.0cm]{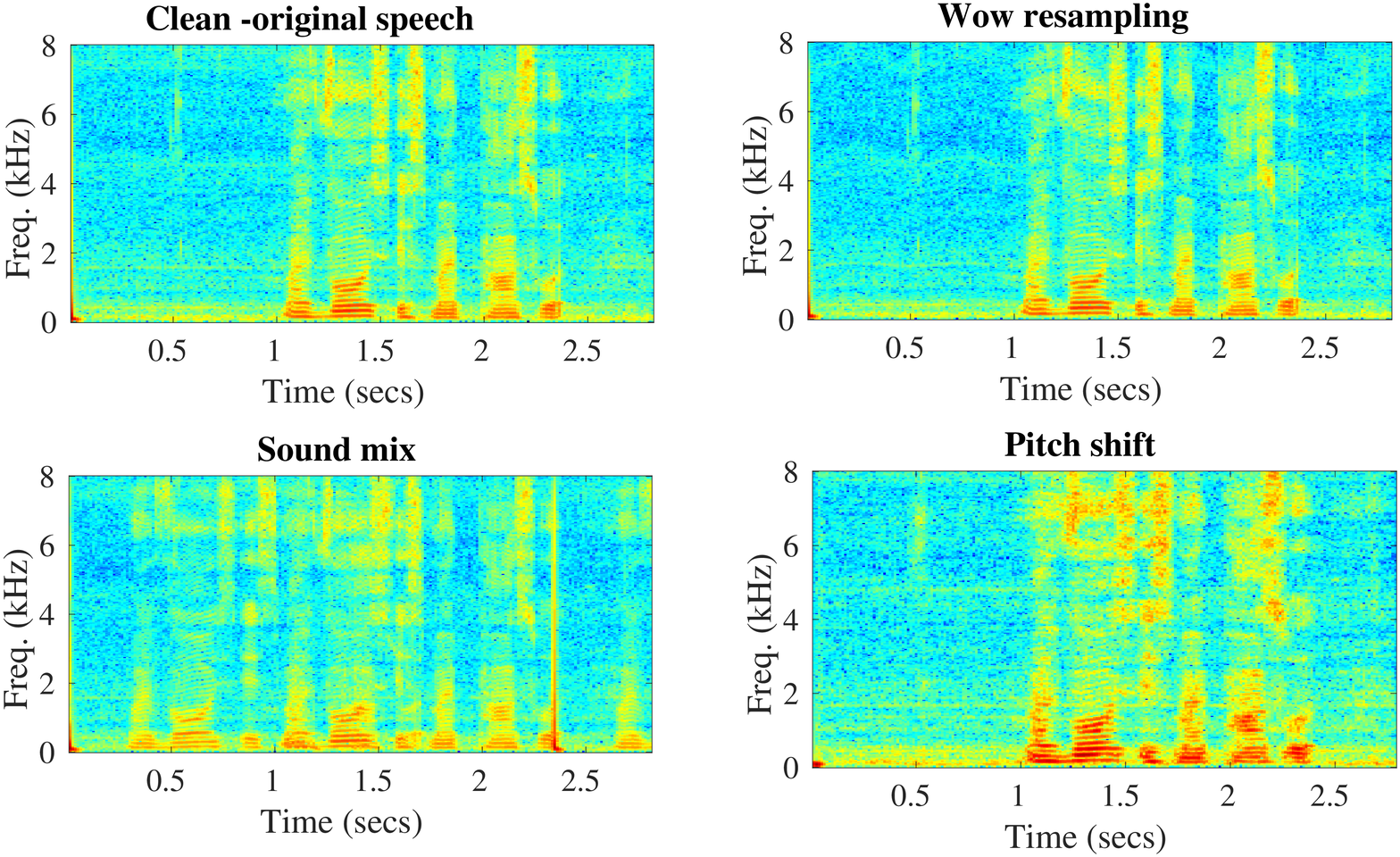} 
\hspace*{-1.90cm}
\includegraphics[width=12.4cm,height=7.0cm]{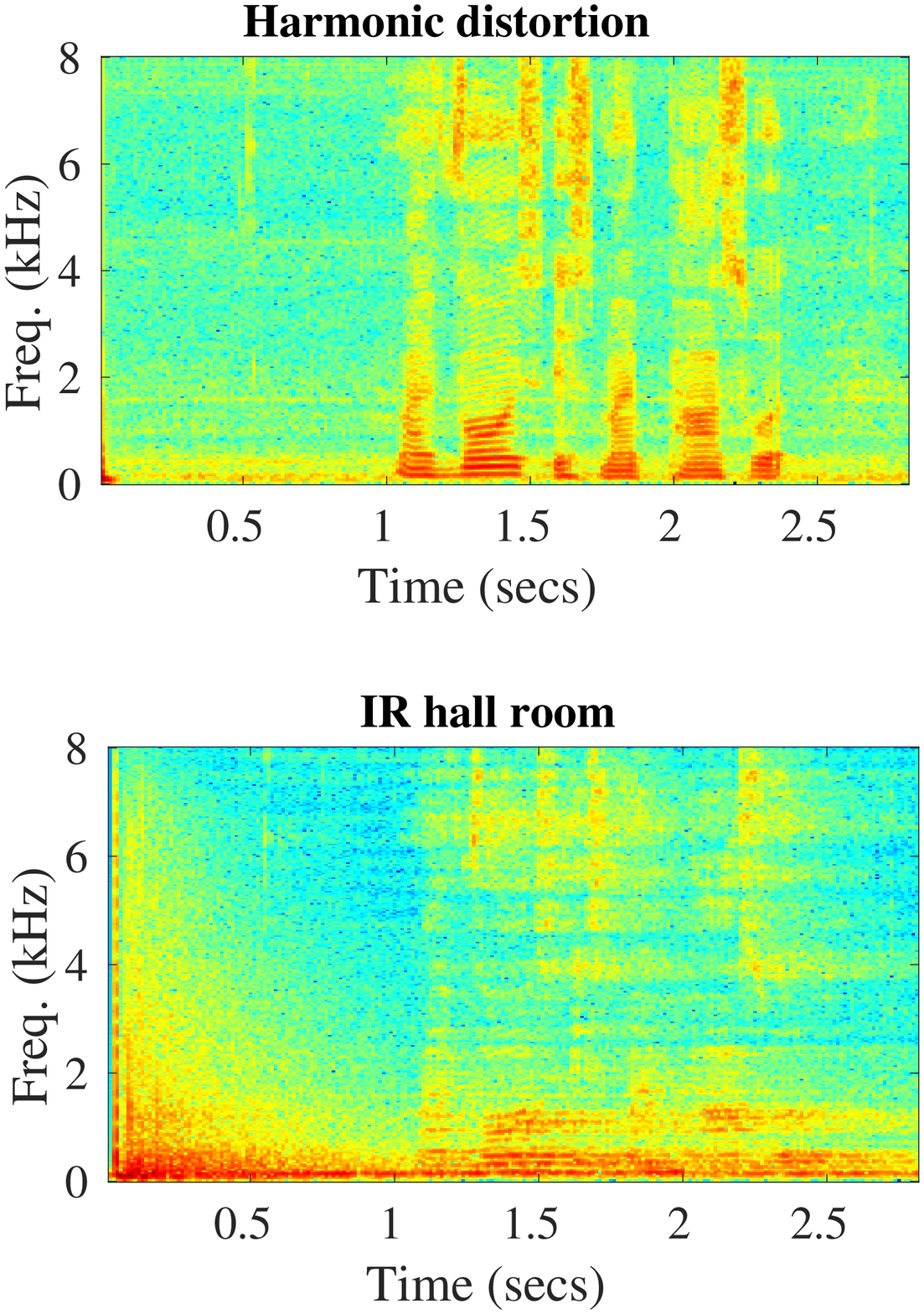} \label{fig:Figs}
\vspace*{-0.80cm}
\caption{\it Spectrograms of the original and the corresponding augmented speech signals. The spoken content in the speech signal is "My voice is my password".}
\label{fig:spectrogram_plot}
\end{figure*}

\section{Experimental setup}
\label{sec:expsetup}
Experiments are performed on the male speakers' parts (task m-part01) of the RedDots challenge 2016 database as per protocol \cite{RedDots}.  There are $320$ target (registered speakers) models  to train and each has three sessions of recording speech samples. Utterances are very short in duration on an average of $2$-$3s$ duration. Database was recorded in the different countries and then send to the other country through different networks to introduce the channel effect in speech signal. More details about the database can be found in \cite{RedDots}.  Four types of test trials are available in the evaluation set to evaluate the system performance in Table \ref{table:no_trial}.

 \begin{itemize}
 \item {\bf Genuine trials:} when a target speaker speaks the pass-phrase/sentence in the test phase, which is same as used during the speaker enrollment phase
\item {\bf Target-wrong:}  when a target speaker speaks a wrong (different) sentence in the testing phase as compared to their enrollment phase
\item {\bf Imposter-correct:} when an imposter speaks the same sentence as that of the target enrollment sessions
\item {\bf Imposter-wrong:} when a imposter speaks a wrong sentence in test phase as compared to the target enrollment pass-phrases
\end{itemize}

For signal processing, $57$ dimensional MFCC \cite{Davis80}(static $C_1$-$C_{19}$, $\Delta,\Delta\Delta$) using $25\; ms$ of hamming window at frame rate of  $10\; ms$. The MFCC features are then processed  with RASTA filtering~\cite{Hermanksy94}. Afterward, robust voice activity detector (rVAD) \cite{TanSD20} algorithm  is applied to discard the less energized frames. Finally,  selected frames are normalized to fit zero mean and unit variance normalization at utterance level.
A gender independent GMM-UBM with $512$ mixtures (having  diagonal co-variance matrices) is trained using $6300$ utterances from ($438$ males, $192$ females) TIMIT database. $3$ iterations are considered during MAP adaptation with the value of relevance factor $10$.  Audio degradation toolbox \cite{Mauch2013} is used to generate the augmentation data. To measure the performance of TD-SV, the equal error rate (EER) and minimum detection cost function (MinDCF) are used as per NIST 2008 SRE \cite{DET97,SRE08}.

\begin{table*}[!ht]
\caption{\it Comparison performance of TD-SV for different enrollment and evaluation data in RedDots database (m-part01 task).}
\vspace*{+0.2cm}
\begin{center}    
\begin{tabular}{llcclll|l}\cline{1-8}
System & Speaker     & \multicolumn{2}{c}{Evaluation}   & \multicolumn{3}{c}{Non-target type [\%EER/(MinDCF$\times$ 100)]}  &  Average\\
       & Enrollment          & Noise & SNR (dB)                  & target-wrong      & imposter-correct & imposter-wrong & EER/MinDCF \\ \hline 
a  (Baseline) &  Original    & - & -                      & 3.96/1.54      & 2.79/1.33    & 0.92/0.25 & 2.55/1.04            \\
                             &   &                        &                &              &           &         \\
                             & & Market  & 5              & 17.39/5.42     & 15.02/4.98   & 11.62/3.12 & 14.68/4.51       \\
                             & &         & 10             & 11.04/3.73     & 9.10/3.42    & 6.04/1.62  & 8.73/2.92  \\
                             &   &                        &                &              &           &         \\
                             & & Car     & 5              &  4.44/1.78     & 3.48/1.69    &1.34/0.40  & 3.09/1.29     \\
                            & &         & 10              &  4.07/1.68     & 3.14/1.60    & 1.20/0.34 & 2.80/1.21       \\ \hline
 b &  Wow resampling         & Market  & 5                & 16.64/5.25     &15.02/5.10    &11.50/3.18  &14.38/4.51        \\
                             & &         & 10                     &    10.79/3.67            &     9.38/3.50       &   5.89/1.66       &8.69/2.95        \\
                             &   &                        &                &              &           &         \\
                             & & Car     & 5              & 4.41/1.70      &    3.79/1.79     &1.57/0.42  &3.25/1.30        \\
                             &&         & 10              & 3.85/1.59      &    3.39/1.68     & 1.30/0.35 & 2.85/1.21   \\   \hline   
 c &  Pitch shift            & Market  & 5                &  29.21/8.84     &26.74/8.87  &    23.59/7.53  &26.51/8.41        \\
                             &&         & 10              &   23.21/7.76             &    20.97/7.80        &     17.48/5.79     &     20.55/7.11   \\
                             &   &                        &                &              &           &         \\
                             && Car     & 5               & 16.89/5.70     &15.05/6.21  &11.53/3.66    &14.49/5.19        \\
                             &&         & 10              &  16.07/5.42    & 14.21/5.96 &10.54/3.38   &13.61/4.92 \\ \hline 
 d &  Harmonic distort       & Market  & 5                &  27.23/7.89     & 25.10/7.90 &    22.86/6.85   &25.07/7.54        \\
                             &&         & 10              &22.41/6.60                & 20.16/6.55           &   17.59/5.35       &   20.05/6.17     \\
                             &   &                        &                &              &           &         \\
                             && Car     & 5               &16.28/5.45      & 13.76/5.12   & 11.19/3.31 & 13.74/4.63       \\
                             &&         & 10              & 15.88/5.36     &13.32/4.92    & 10.30/3.06    &13.17/4.45 \\ \hline 
 e & IR  hall room           & Market  & 5                & 32.47/9.52     &29.61/9.26   &    26.92/8.61    & 29.67/9.13        \\
                             &&         & 10              &    29.61/9.06            &   26.24/8.83         &    23.47/7.93      &  26.44/8.61      \\
                             &   &                        &                &              &           &         \\
                             && Car     & 5               & 24.86/8.32  &21.62/8.12 &18.88/6.57  & 21.79/7.67       \\
                             &&         & 10              & 25.16/8.18   &21.74/8.05     &19.17/6.61    & 22.03/7.61  \\ \hline 
 f &  Sound mix              & Market  & 5                & 22.27/6.58      &18.90/6.26  &16.90/4.82 &    19.36/5.89    \\
                             &&         & 10              &    16.28/5.13            &  13.35/4.75          &   11.32/3.22       &    13.65/4.37    \\
                             &   &                        &                &              &           &         \\
                             && Car     & 5               & 9.44/3.37   &7.52/3.26   &5.33/1.56 &7.43/2.73        \\
                             &&         & 10              & 8.87/3.20   & 7.06/3.13   &4.92/1.46   &6.95/2.60        \\ \hline 
 Score fusion/(a,b,f)    & Maximum& Market  & 5 (dB) &  16.81/5.21 &14.64/5.01    & 10.94/3.05 & {\bf 14.13/4.42}           \\
                         &         &         &   10  &  10.51/3.67 & 9.14/3.44    & 5.95/1.58  & {\bf 8.54/2.90}       \\
                         &         &         &        &                    &                    &                  &        \\
                         &         & Car     & 5  & 4.28/1.73 & 3.54/1.72 & 1.35/0.39 & {\bf 3.06/1.28} \\       
                         &         &         & 10     &3.82/1.62 & 3.17/1.62  & 1.11/0.32 & {\bf 2.70/1.19}   \\ \hline

\end{tabular}
\end{center}
\label{table:GMM-res-reddots_noise}
\end{table*}

\section{Result and discussions}
 \label{sec:resdis}
In this section, we analyze the performance of TD-SV system for different data augmentation methods (in the enrollment phase) and tested on original evaluation  data with or without noise.

\subsection{Effect on TD-SV performance using augmentation data for speaker enrollment when tested on clean  evaluation data}

Table \ref{table:GMM-res-reddots} presents the comparison of TD-SV performance when speaker enrollment is done with or without data  augmentation methods as well as various fusion strategies on the RedDot database (on task m-part01). \emph{Original} indicates the speech files available in the database for training and testing.  It is observed that the \emph{wow re-sampling} augmentation method gives very close values of average EER/MinDCF compared to the baseline, i.e. the conventional system without data augmentation. This indicates that \emph{wow re-sampling} generates content containing most speaker relevant information as compared with other augmentation techniques. 

To further investigate the performance observed above, we plot the spectrograms of an original speech signal and the corresponding augmented data in Fig. \ref{fig:spectrogram_plot}.  From Fig. \ref{fig:spectrogram_plot}, it is noticed that except for the \emph{wow re-sampling} augmentation method, other methods significantly modify the structure of the spectrogram, especially on the higher frequency  components (the most for IR method). These modifications are reflected in the performance of TD-SV. Now if we look at the score fusion among systems, fusion of \emph{a}, \emph{b} and \emph{f} with the \emph{maximum method} yields average EER of $2.46\%$ and MinDCF of $1.04$, both lower than those of the baseline system (\emph{a}). This indicates the effectiveness of augmentation methods. \emph{Multi-condition (a,b,f)} presents the TD-SV with multi-condition training, where a speaker model is derived from GMM-UBM with MAP by pooling original enrollment data along with \emph{wow re-sampling, sound mix} augmentation data. However,  the error rate of the multi-condition training is slightly higher than the baseline. This indicates score fusion is a better choice for TD-SV using augmentation data.

\subsection{Effect on TD-SV performance using augmentation data for speaker enrollment when tested on a noisy version of the evaluation data}
Experimental results on the noisy version of the evaluation data are presented in Table \ref{table:GMM-res-reddots_noise}. The noisy version of the evaluation data is generated as per ITU protocol \cite{ITU2005}. For simplicity, we  present only the system performance for  \emph{market} and \emph{car noise scenarios} for SNR values of $5$ and $10$ dB, and maximum method (found optimal in previous subsection) in score fusion. 
The multi-condition method is not studied here as it does not improve the SD-SV as shown in Table \ref{table:GMM-res-reddots}.   From Table \ref{table:GMM-res-reddots_noise}, it can be seen that error rates of all systems  significantly increase and are expected due to the mismatch between the enrollment with clean data and the evaluation with noisy data. Similarly to the Table \ref{table:GMM-res-reddots},  \emph{wow re-sampling} and \emph{Sound mix} show lower error rates than the other augmentation methods and fusion further improves the TD-SV with respect to the baseline.  This indicates the usefulness of the data augmentation in TD-SV under noisy conditions.

\section{Conclusion}
\label{sec:con}
In this paper, we proposed to explore a set of data augmentation approaches for generating extra data to empower speaker enrollment, in contrast to the use of data augmentation for building speaker independent models for TD-SV, in low resource applications. In the proposed method, each speaker is represented by a number of models that are derived from GMM-UBM using the original enrollment data together with augmented data (generated from the original enrollment data) for the particular speaker. It gives different TD-SV systems corresponding to different augmentation approaches. In the test, a test utterance is scored against different systems and then  fused them with different strategies. Besides, we also evaluated the performance of TD-SV under clean and noisy environment conditions. Experimental results depicted that score fusion of the conventional/baseline system with the proposed data augmentation system reduces the error rate of TD-SV compared to their standalone counterpart. Experiments were conducted on the RedDots challenge 2016 database.


 \bibliographystyle{IEEEbib}
 \bibliography{strings,References}
 \end{document}